\documentclass[prb,twocolumn,showpacs,superscriptaddress]{revtex4}
\usepackage[dvips]{epsfig}
\usepackage{graphicx}
\usepackage{amsmath,amssymb}
\usepackage{bm}
\usepackage{bbm}
\usepackage{dsfont}
\newcounter{fignum}

\newcommand{\hc}{\mathop{\rm H.c.}\nolimits}
\newcommand{\ket}[1]{{\vert #1 \rangle}}
\newcommand{\bra}[1]{{\langle #1 \vert}}
\newcommand{\dg}{\dagger}

\newcommand{\skalp}[2]{\langle #1  \vert #2 \rangle}

\newcommand{\ga}{{\alpha}}

\newcommand{\gd}{{\delta}}

\newcommand{\gpsi}{{\psi}}

\newcommand{\gphi}{{\phi}}
\newcommand{\gD}{{\Delta}}
\newcommand{\gr}{{\rho}}
\newcommand{\gvphi}{{\varphi}}

\newcommand{\hH}{{\hat H}}
\newcommand{\mbf}{\mathbf}
\renewcommand{\vec}{\mbf}
\newcommand{\veck}{\vec k}
\newcommand{\vecr}{\vec r}

\newcommand{\cE}{{\cal E}}
\begin{document}

\title{Character of electronic states in graphene antidot lattices:\\
       Flat bands and spatial localization}
\author{Mihajlo Vanevi\'c}
\affiliation{School of Physics, Georgia Institute of Technology, Atlanta,
             Georgia 30332, USA}

\author{Vladimir M. Stojanovi\'c}
\affiliation{Department of Physics,
             Carnegie Mellon University, Pittsburgh, Pennsylvania 15213, USA}
\affiliation{Department of Physics, University of Basel,
 Klingelbergstrasse 82, CH-4056 Basel, Switzerland}

\author{Markus Kindermann}
\affiliation{School of Physics, Georgia Institute of Technology, Atlanta,
             Georgia 30332, USA}

\date{\today}

\begin{abstract}
Graphene antidot lattices have recently been proposed as a new breed of graphene-based
superlattice structures. We study electronic properties of triangular antidot lattices, with
emphasis on the occurrence of dispersionless (flat) bands and the ensuing electron
localization. Apart from strictly flat bands at zero energy (Fermi level), whose existence is
closely related to the bipartite lattice structure, we also find quasi-flat bands at low energies.
We predict the real-space electron density profiles due to these localized states for a
number of representative antidot lattices. We point out that the studied low-energy, localized
states compete with states induced by the superlattice-scale defects in this system which have
been proposed as hosts for electron spin qubits. Furthermore, we suggest that local moments formed in
these midgap zero-energy states may be at the origin of a surprising saturation of the electron
dephasing length observed in recent weak localization measurements in graphene antidot lattices.
\end{abstract}
 \maketitle

%73.43.-f Quantum Hall effects
%73.21.Cd Superlattices
%% 71.70.Di Landau levels

\section{Introduction}

Investigations of the electronic properties of graphene constitute a relatively
new and thriving sub-area of condensed-matter research.~\cite{GrapheneReview}
After the first successful fabrication of graphene monolayers~\cite{Novoselov:04}
-- by means of a mechanical exfoliation of graphite -- this two-dimensional
semimetallic material has ignited tremendous experimental~\cite{Zhou:07,Wu+:07,Zhou:08} and
theoretical~\cite{Tworzydlo:06,Peres+:06,EdgePotential,Cheianov:06,Katsnelson+:06,Hwang+DasSarma,Trauzettel:07,Milton+:07,Recher:07,Stauber:08}
interests. From a fundamental standpoint, one of the main incentives for studying graphene stems
from emergent analogies between the low-energy physics of the material and relativistic
quantum mechanics.~\cite{Semenoff+DiVincenzoMele,BeenakkerRMP:08,Stander:09}
Graphene is just as interesting from a practical point of view: owing to its exceptional
properties -- the extremely high mobility, chemical inertness, atomic thickness, and easy control of
charge carriers by applied gate voltages -- it holds a great promise for a
carbon-based ``post-silicon'' microelectronics.~\cite{GrapheneDevices,Wang+:08,IBMtransistor:09,Avouris+:07}

Aside from simple graphene monolayers, patterning
of monolayer films by nanolithography methods~\cite{Berger:06}
-- allowing feature sizes as small as tens of nanometers --
has led to the demonstration of nanostructures such as Hall
bars,~\cite{Molitor+:07,Giesbers+:08,Ki+:08}
quantum dots,~\cite{Geim+Novoselov:07} nanoribbons,~\cite{Han:07,IBMribbon:08,Stampfer+:09}
and Aharonov-Bohm interferometers.~\cite{Russo+:08}
Another family of graphene-based structures has recently been proposed --- triangular
superlattices of holes (antidots) cut in a graphene sheet, known as antidot
lattices.~\cite{PedersenAntidot:08,Pedersen+:08} Unlike pristine graphene which is semimetallic,
graphene antidot lattices are semiconducting, with a direct band gap that depends on the antidot
size. Square antidot lattices have been studied experimentally quite recently, corroborating the
existence of a transport gap.~\cite{Shen:08pre,Eroms:09pre} In addition, the weak localization
correction to the conductance and a surprising saturation of the electron dephasing length at
the superlattice scale have been observed in these experiments.

In a recent work,~\cite{PedersenAntidot:08} antidot lattices have been proposed as
a platform for quantum computation, with defects in this system envisioned as
hosts for electron spin qubits. While the proposal of Ref.~\onlinecite{PedersenAntidot:08}
focuses on localized states due to defects on the {\em superlattice scale} (such as missing antidots),
their counterparts on the {\em lattice scale} (such as vacancies or adatoms),
essentially unavoidable along the antidot edges of experimental samples, are known to
give rise to midgap (bound) states as well.~\cite{Yazyev:08,RossierPalacios} One may thus expect that the
two types of defects compete. In addition, midgap states caused by lattice-scale defects
might provide a plausible explanation of the maximal dephasing length observed in the experiment of
Ref.~\onlinecite{Eroms:09pre}: for sufficiently large charging energies, such midgap states can host
local (spin) moments at the antidot edges that are known to be a very effective source of electron
dephasing.~\cite{dephasing} This motivates us to systematically study midgap states in graphene antidot
lattices, with a focus on their spatial profile and the corresponding dispersionless (flat) bands.

On bipartite lattices (such as graphene) which have an excess of atoms on one of the two sublattices, zero-energy
states are expected on very general grounds.~\cite{Lieb:89,Inui:94} As a consequence, midgap states
may exist even in perfectly symmetric and periodic antidot lattices.
This was put forward by Shima and Aoki~\cite{Shima+Aoki:93} in a general symmetry-based classification
of superhoneycomb systems (i.e., triangular superlattices based on an underlying honeycomb lattice).
Generically, however, midgap states will be introduced by
irregularities in the shape of the antidot lattice on the atomic scale, which are unavoidable in
present-day experiments. We analyze the resulting band structure and the spatial profile of the corresponding
wave functions for a number of representative antidot lattices. Furthermore, we determine  low-energy
tunneling current distributions that can be compared with scanning tunneling microscopy (STM) measurements.
Our estimates for the charging energies of the predicted zero-energy states indeed suggest that for typical
experimental parameters such states can form local moments and thus provide strong dephasing upon
electron scattering from the antidot edges.

The remainder of the paper is organized as follows.
In Sec.~\ref{sectwo}, we first briefly introduce the graphene superlattices of
interest, accompanied by the notation and conventions
to be used throughout (Sec.~\ref{antidotlatt}), and then lay out the framework for
calculating their band structure (Sec.~\ref{bandstruct}) and the tunneling current
distribution (Sec.~\ref{tunnelcurrent}). The obtained results are presented and
discussed in view of the generic properties of the superhoneycomb- and bipartite systems
in Sec.~\ref{resdisc}. Finally, we summarize our findings and conclude in Sec.~\ref{concl}.

\section{Description of antidot lattices}  \label{sectwo}
\subsection{Structure and nomenclature} \label{antidotlatt}

\begin{figure}[t!]
\begin{center}
\includegraphics[scale=0.58]{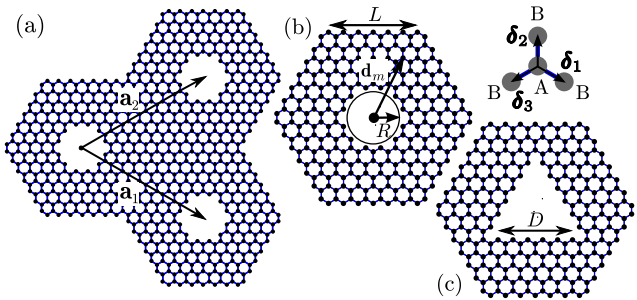}
\caption{\label{antidotab} Graphene antidot lattice: (a) Lattice structure described by the
basis vectors $\mathbf{a}_{1}$ and $\mathbf{a}_{2}$, with magnitude \:$|\mathbf{a}_{1}|=|\mathbf{a}_{2}|
=La\sqrt{3}$\:; (b) hexagonal unit cell with a circular antidot; and (c) hexagonal unit cell
with a triangular antidot. Vectors $\bm{\delta}_{1},\:\bm{\delta}_{2},\:\bm{\delta}_{3}$
specify positions of the nearest neighbors of a carbon atom on sublattice $A$.}
\end{center}
\end{figure}

To set the stage, in this section we introduce graphene antidot lattices.
A segment of a typical antidot lattice with a circular perforation is depicted
in Fig.~\ref{antidotab}(a), with lattice basis vectors denoted by
$\mathbf{a}_{1}$ and $\mathbf{a}_{2}$. Its unit cell is a hexagon with an antidot in
the center [Fig.~\ref{antidotab}(b)]. We characterize the structure by
the dimensionless side length of the
hexagonal unit cell ($L$) and the radius of the antidot ($R$), both
expressed in units of the graphene lattice constant $a=2.46\:$\AA.
(Note that while $L$ is an integer, $R$ can also take non-integer values;
$a=a_{cc}\sqrt{3}$, where $a_{cc}=1.42\:$\AA\: is the distance
between nearest-neighbor carbon atoms.) Therefore, we use the notation $\{L,R\}$
to specify antidot lattices with circular perforations.
The unit cell of an antidot lattice with a triangular perforation,
which can be characterized by $\{L,D\}$ (where $D$ is the side length of the triangle),
is shown in Fig.~\ref{antidotab}(c). The number of carbon atoms (hereafter C atoms)
per unit cell of an antidot lattice will henceforth be denoted by $N_{\textrm{C}}$.

If we take a C atom on sublattice A at the origin, its
nearest neighbors are determined by vectors $\bm{\delta}_{1}=(\sqrt{3}/2,-1/2)\:a_{cc}$,
$\bm{\delta}_{2}=(0,1)\:a_{cc}$, and $\bm{\delta}_{3}=(-\sqrt{3}/2,-1/2)\:a_{cc}$
(see Fig.~\ref{antidotab}).
Alternatively, for a C atom on sublattice B   at origin, the corresponding
vectors are $-\bm{\delta}_{1}$,$-\bm{\delta}_{2}$, and $-\bm{\delta}_{3}$.

\subsection{Method for band-structure calculation} \label{bandstruct}

Given the large size of the unit cells in our superlattice -- that in the
cases of practical interest, with   antidot diameter larger than about $10$\:nm,
have numbers of C atoms in excess of several thousands -- a band-structure calculation using
the standard {\em ab-initio} methods based on the density functional theory (DFT)
is not feasible. We thus compute the electronic band structure of graphene antidot lattices
within a $\pi$-orbital tight-binding model. This  method is known
to reproduce very accurately the low-energy
part of the DFT band structure of pristine graphene.~\cite{Reich:02}

The tight-binding Hamiltonian of an antidot lattice reads
\begin{equation}\label{}
\hat{H}_{\textrm{e}}=-\frac{t}{2}\sum_{\mathbf{R},m,\bm{\delta}}
\big(\hat{a}^\dg_{\mathbf{R}+\mathbf{d}_{m}+\bm{\delta}}
\hat{a}_{\mathbf{R}+\mathbf{d}_{m}}+\hc\big) \:,
\end{equation}
\noindent where vectors $\mathbf{R}$ designate unit cells
(in total $N$ of them), $\mathbf{d}_{m}$ ($m=1,\ldots,N_{\textrm{C}}$)
specify positions of C atoms within a unit cell, $\bm{\delta}$ stands for
nearest neighbors of a C atom at position $\mathbf{R}+\mathbf{d}_{m}$,
and $t\approx 2.8$ eV is the nearest-neighbor hopping matrix element,
while the factor $1/2$ is needed to correct for double-counting.
After Fourier transformation to momentum-space, the band-structure
calculation amounts to a sparse-matrix diagonalization problem of
dimension $N_{\textrm{C}}$.

The Bloch wave functions corresponding to the energy eigenvalues
$\cE_{n\ga}(\veck)$ are given by
$\gpsi_{n\mbf k\ga}(\mbf r)=\sum_m C_m^{(n\veck\ga)}\gphi_{m\veck}(\vecr)$,
where $\gphi_{m\veck}(\vecr)=N^{-1/2} \sum_{\mbf R}
e^{i\veck \cdot \mbf R}\gvphi(\vecr - \mbf R - \mbf d_m)$.
Here $n$ enumerates energy bands (with possible additional degeneracy
labelled by $\ga$) and $\gvphi(\vec r)$ denotes the $2p_z$ orbital of a
C atom. The energy eigenvalue problem
$\hH \ket{\gpsi_\veck} = \cE \ket{\gpsi_\veck}$ reduces to
$\det\left [\mbf H(\veck) - \cE \mbf S(\veck)\right] = 0$,
with matrices $\mbf H$ and $\mbf S$ given by $H_{mm'}(\veck)
=\bra{\gphi_{m\veck}}\hH\ket{\gphi_{m'\veck}}$ and
$S_{mm'}(\veck)=\skalp{\gphi_{m\veck}}{\gphi_{m'\veck}}$,
respectively. In the nearest-neighbor approximation
%%%%%%%%%%%%%%%%%%%%%%%%%%%%%%%%%%%%%%%%%%%%%%%%%%%%%%%%%%%%%%%%%%%%%%%%%%%%
\begin{equation}
H_{mm'}(\veck)=
-t \bigg( \gd_{\langle \vec d_m, \vec d_{m'} \rangle}
+ \sum_{\vec R} e^{i\vec k \cdot \vec R}
\gd_{\langle \vec d_m, \vec d_{m'} + \vec R \rangle}
\bigg) \:,
\end{equation}
\noindent where the summation runs over superlattice vectors
$ \vec R=\pm\vec a_1$, $\pm \vec a_2$, $\pm(\vec a_1-\vec a_2)$
(only neighboring unit cells contribute), and
$\gd_{\langle \vec r, \vec r' \rangle}=1$ if C atoms
at positions $\vec r$ and $\vec r'$ are nearest neighbors,
while $\gd_{\langle \vec r, \vec r' \rangle}=0$ otherwise.
To a good approximation, the overlap of $2p_z$ orbitals on
different C atoms can be neglected, so that $S_{mm'}(\vec k) = \gd_{mm'}$.
This is a standard practice in the analyses of $\pi$-electron
systems.~\cite{hannewald:04}

For later reference, we describe a construction of an orthonormal
eigenbasis $\{\gpsi_{n \vec k \ga}\vec  \}$
of $\hH$. To that end, we first note that
$\gphi_{m\vec k}$ form an orthonormal set:
$\skalp{\gphi_{m\vec k}}{\gphi_{m'\vec k'}} = \gd_{mm'}
\gd_{\vec k \vec k'}$.
[For $\vec k = \vec k'$ the orthogonality follows from
$S_{mm'}(\vec k) = \gd_{mm'}$, while for $\vec k \neq \vec k'$
it holds because $\gphi_{m\vec k}$ and $\gphi_{m'\vec k'}$
belong to different eigensubspaces of the lattice-translation
operator.] This implies that
$\skalp{\gpsi_{n\vec k\ga}}{\gpsi_{n'\vec k'\ga'}} = 0$
for $\vec k \neq \vec k'$. Therefore, to construct an orthonormal set,
we find an orthonormal basis in the degenerate eigensubspaces
of $\mbf{H}(\mbf{k})$. The orthogonality of the latter basis
$\sum_m (C_m^{(n\vec k\ga)})^* C_m^{(n\vec k\ga')}= \gd_{\ga\ga'}$
implies that $\skalp{\gpsi_{n\vec k\ga}}{\gpsi_{n\vec k'\ga'}}
= \gd_{\vec k\vec k'} \gd_{\ga\ga'}$.

\subsection{Tunneling current distribution} \label{tunnelcurrent}

STM and related techniques are known as powerful tools for mapping out the
spatial form of  surface electron states.~\cite{Binnig+Rohrer:99}
In order to visualize the spatial structure of the studied low-energy states
and provide the means of comparing our results to STM measurements,
we predict the tunneling current distribution.

The tunneling current
\begin{equation}
{\cal I}(\vec r) \propto \int_{\cE_F}^{\cE_F + eV}
d\cE \; \gr(\vec r; \cE),
\label{eq:IofR}
\end{equation}
with $\cE_F$ being the Fermi energy and $V$ the STM-tip bias voltage,
can be used to probe the spatial
dependence of the local density of states~\cite{GrossoParravicini}
\begin{equation}
\gr(\vec r; \cE)=\sum_{n\vec k\ga}|\gpsi_{n\vec k\ga}(\vec r)|^2\;
\gd[\cE - \cE_n(\vec k)] \:.
\end{equation}
\noindent This is a basis-independent quantity, expressed here in terms of
an {\it orthonormal} eigenbasis $\{\gpsi_{n\vec k\ga} \}$ of
$\hat{H}$, constructed as described above. In the case that only one flat
band at $\cE=\cE_{n_0}$ falls into the energy window in Eq.~\eqref{eq:IofR}, we have
\begin{multline}\label{eq:I}
{\cal I} (\vec r) \propto \sum_{\vec k \ga} |\gpsi_{n_0\vec k \ga}(\vec r)|^2 \\
=\sum_{\vec k\ga} \sum_{mm'}
(C_m^{(n_0\vec k\ga)})^* C_{m'}^{(n_0\vec k\ga)}\;
\gphi_{m\vec k}^*(\vec r) \gphi_{m'\vec k}(\vec r) \:.
\end{multline}

Strictly speaking, to calculate the tunneling current we would need to
use the explicit form of $2p_z$ orbitals. However, Eq.~\eqref{eq:I} simplifies
if we assume that the $2p_z$ orbitals are well-localized on C atoms
and hence neglect the overlap of the neighboring orbitals. After coarse-graining
$I({\vec r})= (1/\gD V)\int_{\gD V\ni \vec r} d^3\vec r'\; {\cal I}(\vec r')$
in the vicinity of a C atom positioned at
$\vec r = \vec R+\vec d_m$, we obtain the sought-after tunneling current
distribution
\begin{equation}
I({\vec r})\propto \sum_{\vec k\ga} |C_{m}^{(n_0\vec k\ga)}|^2 \:,
\end{equation}
\noindent which is a lattice-periodic quantity.

\section{Results and Discussion} \label{resdisc}

\begin{figure}[t!]
\begin{center}
\includegraphics[scale=0.7]{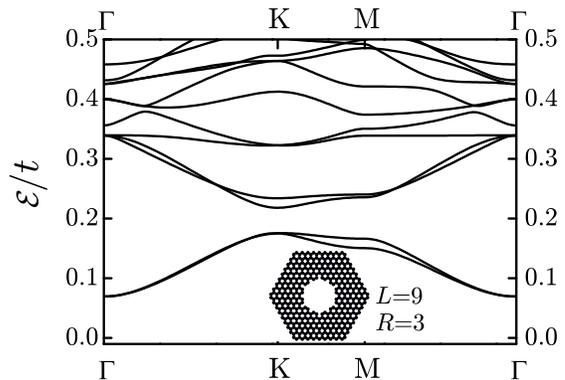}
\caption{\label{fig:Ekband93lattice}
Band structure for a \{9,3\} antidot lattice. Only bands
above the Fermi level ($\cE=0$) are shown because of the
particle-hole symmetry. $\Gamma$, K, M  stand for the
high-symmetry points in the Brillouin zone.}
\end{center}
\end{figure}

In this section, we investigate the band structure
of antidot lattices, with emphasis on localized states
at low energy. The spatial profile of these states
is characterized by a tunneling current distribution.
We present results for both ideal lattices
and lattices with defects (vacancies and/or debonded C atoms).
Since we describe the system by a nearest-neighbor
tight-binding model on a bipartite lattice, the resulting energy
spectrum has  particle-hole symmetry.~\cite{FazekasBook}
With this in mind, in the following we focus  on the bands
above or at the Fermi level ($\cE=0$).

We start with an antidot lattice that is made up of
perfect circular perforations, such as those considered in
Ref.~\onlinecite{PedersenAntidot:08}.
The band-structure of an $\{L=9,R=3\}$ antidot lattice is shown in
Fig.~\ref{fig:Ekband93lattice}. A band gap $\cE_{\textrm{g}}$ opens at
the $\Gamma$-point ($\mathbf{k}=0$).
For antidot lattices with a relatively small number of removed C atoms
($N_{\textrm{rem}}$) compared to the total number of atoms in the
original unit cell ($N_{\textrm{total}}=N_{\textrm{rem}}+N_{C}$),
the band gap has been predicted to scale as~\cite{PedersenAntidot:08}
$\cE_{\textrm{g}}\propto \sqrt{N_{\textrm{rem}}}/N_{\textrm{total}}$.
% This band gap can be tuned by engineering an antidot lattice with suitably
% chosen $L$ and $R$.
It has been demonstrated that localized states with
energy inside the gap can be induced in this system by defects on the
{\em superlattice scale}, e.g., missing circular antidots in the
lattice.~\cite{PedersenAntidot:08}
Such localized states have been proposed as hosts for local spin moments that may
be utilized for quantum computation. In the following, we show that these
midgap states appear generically in graphene antidot lattices, even without
superlattice-scale defects.

In the nearest-neighbor approximation, graphene has a bipartite lattice, that is,
a lattice that can be divided into two sublattices, A and B, where only sites at
different sublattices are connected through nonzero hopping matrix elements.
Inui {\em et al.}~\cite{Inui:94} showed that in such systems
one has $N_{A}-N_{B}$ zero-energy states, where $N_{A},\:N_{B}$
are the total number of sites on the respective sublattices.
[This result was, in fact, implicitly known even earlier: it was obtained by
Lieb~\cite{Lieb:89} as a prerequisite for the proof that the total spin in the
exact ferromagnetic ground state of the Hubbard model on a bipartite lattice
is $S=(N_{A}-N_{B})/2$.] In a graphene superlattice with $n_{A}$ and $n_{B}$ sites
per unit cell on the A- and B-sublattices, respectively,
these states form $n_{A}-n_{B}$ dispersionless (flat) bands at $\cE=0$. A sublattice
imbalance $n_{A}-n_{B}$ is generically introduced at the edges of graphene-based
structures. Perhaps the most well-known
example of the corresponding zero energy states are the edge states in
zigzag graphene ribbons that form a {\em partially flat} band at the Dirac
point.~\cite{Fujita:96,Nakada+:96} A pair of essentially flat (spin-polarized) bands
close to the Fermi level was found also in a hydrogenated graphene ribbon, as
demonstrated by {\em ab-initio} electronic structure calculations.~\cite{Kusakabe:03}

\begin{figure}[t!]
\begin{center}
\includegraphics[scale=1.9]{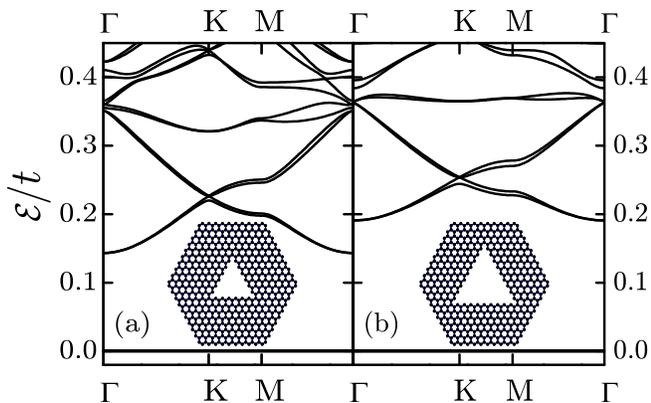}
\caption{\label{fig:antidotTriangleL10R3L10R5}
Band structure of an antidot lattice with triangular antidots.
Only bands above the Fermi level ($\cE=0$) are shown because of the
particle-hole symmetry. $\Gamma$, K, M  stand for the high-symmetry points in
the Brillouin zone. The flat band at $\cE=0$ is (a) sixfold degenerate ($D=6$),
and (b) ninefold degenerate ($D=9$).}
\end{center}%
\vspace*{-3mm}
\end{figure}

In antidot lattices one expects flat bands at zero energy due to
sublattice imbalances along the edges of the antidots. The sublattice imbalance
can occur even for perfect regularly shaped antidots. As an example, we study antidot
lattices with triangular perforations (cf. Fig.~\ref{antidotab}), which invariably
have an imbalance of $n_{A}-n_{B}=D$ per antidot. Consequently, a $D$-fold degenerate
flat band emerges at $\cE=0$, as depicted in Fig.~\ref{fig:antidotTriangleL10R3L10R5}.
On the other hand, it is easy to check that  antidot lattices with perfect
circular perforations always have $n_{A}=n_{B}$. These lattices therefore do not
exhibit flat bands at $\cE=0$ (cf. Fig.~\ref{fig:Ekband93lattice}).

\begin{figure}
\begin{center}
\includegraphics[scale=1.6]{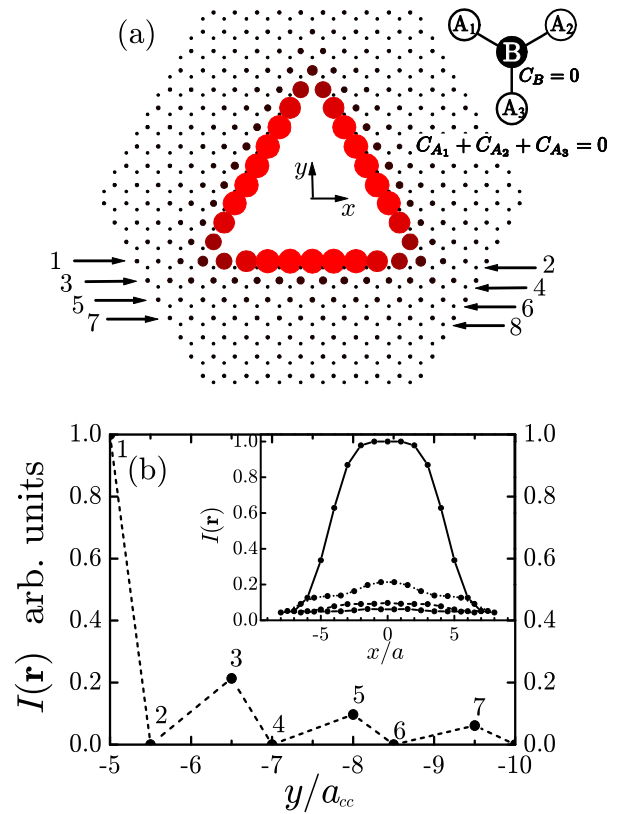}
\caption{\label{fig:IofRtridotLL10RR5} Illustration of the spatial localization
of the single-particle wave function corresponding to a flat band.
(a) Tunneling current distribution $I({\vec r})$,
reflecting the population of only the A-sublattice sites. The sizes of the circles and
the color brightness are proportional to $I({\vec r})$, which is scaled to
$0\le I \le 1$. The electron amplitudes obey the sum rule shown in the inset.
(b) Tunneling current maxima along the $y$ direction,
at positions indicated by the arrows in (a). Inset: tunneling current
distribution in the $x$ direction. }
\end{center}
\end{figure}

As pointed out by Inui {\em et al.},\cite{Inui:94} the single-particle states corresponding
to the zero energy flat bands are pseudospin-polarized -- they occupy only sites belonging to a
particular sublattice. Figure~\ref{fig:IofRtridotLL10RR5} shows for an antidot lattice
$\{L=10,D=9\}$ how this effect manifests itself in the tunneling current distribution, which is
proportional to the on-site electron density. Closer inspection reveals that
the single-particle states $\psi_{\mathbf{k}}$ corresponding to the flat band at $\cE=0$
indeed leave the B-sublattice sites unoccupied (amplitudes $C_{B_m}=0$),
while the A-sublattice sites are occupied with amplitudes $C_{A_m}$ (normalized as
$\sum_{m}|C_{A_m}|^{2}=1$), adding up to zero around each B-sublattice site
[see Fig.~\ref{fig:IofRtridotLL10RR5}(a)]. Moreover, the electronic states corresponding
to the zero-energy flat bands are predominantly localized in the vicinity of the antidot
edge.~\cite{Jafri+:09pre}
The extent of their spatial localization is illustrated in Fig.~\ref{fig:IofRtridotLL10RR5}(b).

It is important to emphasize that the above results are valid for an arbitrary
system of bipartite structure, either infinite or of finite size, and even in the presence
of {\em off-diagonal} disorder and/or an   external magnetic field.~\cite{Inui:94}
(In the presence of a magnetic field, the hopping matrix elements become complex through
the conventional Peierls substitution~\cite{Peierls:33} and the particle-hole symmetry
is destroyed.) Therefore, even without actual calculation,
we can conclude that the discussed flat bands at $\cE=0$ in graphene antidot lattices
remain flat in a  magnetic field. Analogous results for finite-size graphene
antidot flakes have recently been obtained numerically.~\cite{Bahamon+:09}

A different perspective on the problem is furnished by a symmetry-based classification of
superhoneycomb systems, as put forward by Shima and Aoki.~\cite{Shima+Aoki:93} They showed
that such systems can have either semiconducting (with direct band gap), semimetallic, or
metallic character. This classification turns out to depend not only on the global superlattice symmetry, but
also on the specific atomic configuration within the unit cell. In particular, structures with
$6m$, $6m+2$, $6m+3$, and $6m+5$ ($m$ is an integer) atoms per unit cell belong to  symmetry classes
designated by $A_{0}$, $A_{C}$, $B_{0}$, and $B_{C}$, respectively.~\cite{Shima+Aoki:93}
The respective degeneracies of the flat bands at $\cE=0$ in these classes are
$6l$, $6l\pm 2$, $6l-3$, and $6l\pm 1$ ($l\geq 0$ is an integer).
According to this classification, our antidot lattices with circular perforations
belong to the $A_{0}$-type of superhoneycomb systems with $l=0$, which is consistent
with the absence of flat bands at $\cE=0$.
\begin{figure}
\includegraphics[scale=2.39]{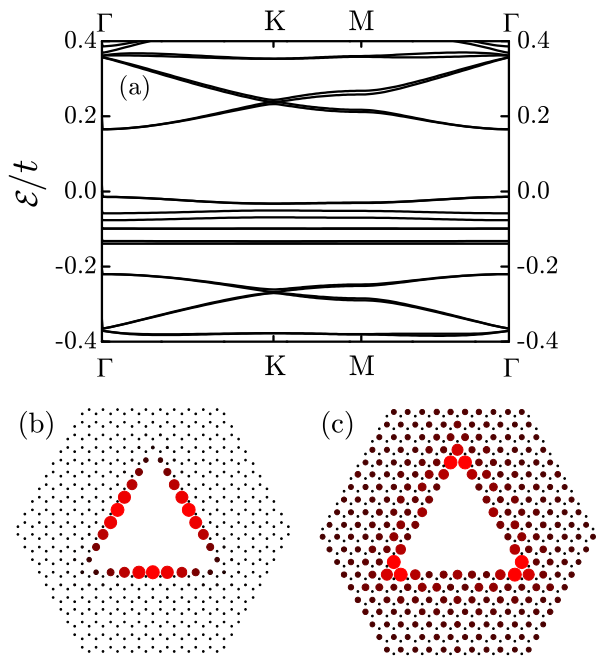}
\caption{\label{fig:bandstructV01}
Effect of an on-site impurity potential $V_{e}=-0.15\:t$ along the
antidot edges:
(a) band structure, and the tunneling current distributions corresponding to
(b) the lowest-lying flat band and (c) the highest flat band.}
\end{figure}%

In realistic graphene flakes, the dangling bonds along the edges
are hydrogen-passivated, giving rise to an on-site potential.
An on-site potential can also appear in a pure carbon system, due to
a weak edge-magnetization induced by electron-electron interactions.~\cite{EdgePotential}
Since the electronic states corresponding to the zero-energy flat
bands are largely localized at the antidot edges, one
expects them to be strongly affected by such an edge potential.
This motivates us to study the influence of an edge potential, hereafter
denoted by $V_{e}$, on the zero-energy states. We assume that
$V_e$ takes nonzero values at C atoms that have only two nearest neighbors.
Such a potential breaks the particle-hole symmetry of the energy spectrum by destroying
the bipartite lattice structure, since it couples sites that belong to the same sublattice.
However, the main effect of $V_{e}$ is a lifting of the degeneracy of the
zero-energy flat bands, while their dispersionless character is largely being
preserved.~\cite{commentLL} At the same time, the effect of this potential
on the other bands is relatively small [cf. Figs.~\ref{fig:bandstructV01}(a) and
\ref{fig:antidotTriangleL10R3L10R5}(b)]. This finding is indeed analogous to
the behavior of flat bands in nanoribbons with hydrogenated edges, obtained using
realistic first-principles DFT calculations.~\cite{Kusakabe:03} We stress that the
observed effect of the on-site potential is robust, i.e., changing the magnitude of
$V_{e}$ does not alter our qualitative conclusions.

\begin{figure}[t!]
\begin{center}
\includegraphics[scale=2.39]{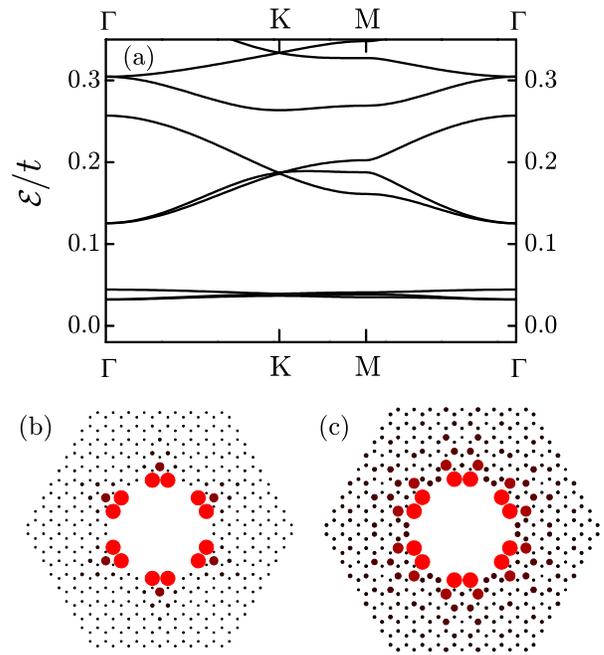}
\caption{\label{fig:IofRcircleDanglingBondsL9R3d2ab}
Effects of debonded C atoms in the antidot lattice $\{9,3.2\}$:
(a) band structure, and the tunneling current distributions corresponding to
(b) the low-lying quasi-flat bands and (c) the energy bands
within the range $0.125\:t\lesssim \cE \lesssim 0.257\:t$.}
\end{center}
\end{figure}

In what follows, we point out another generic feature of graphene antidot lattices:
the occurrence of essentially flat bands at nonzero energies (even in the absence of an
edge potential, $V_e=0$). A characteristic example is presented
in Fig.~\ref{fig:IofRcircleDanglingBondsL9R3d2ab}.
While flat bands at $\cE=0$ in bipartite lattices arise due to a
global sublattice imbalance ($n_{A}\neq n_{B}$), as discussed above, we ascribe
the quasi-flat ones at $\cE \neq 0$ to local sublattice imbalances (while globally $n_{A}=n_{B}$).
Such local imbalances can be induced even in regularly shaped antidot lattices,
for instance, by debonded C atoms with a single neighbor at the edges
[cf. Fig.~\ref{fig:IofRcircleDanglingBondsL9R3d2ab}(b)].

The occurrence of quasi-flat bands may be understood as follows.
A single debonded C atom induces a sublattice imbalance that leads
to a zero energy  ``defect level''.
One may view a collection of debonded C atoms as a collection of ``local sublattice
imbalances'' that induce one defect level each  with
wave functions that are localized in the vicinity of the defects.
This picture is supported by the tunneling
density of states shown in Fig.~\ref{fig:IofRcircleDanglingBondsL9R3d2ab}(b).
Localized states induced by defects that are
well separated from one another
hybridize only weakly. Accordingly, these defect levels give rise to
essentially dispersionless bands close to the Fermi level ($\cE=0$).~\cite{JpnFlatBands}
As the distance between the defects decreases,
the hybridization of defect states becomes stronger
and the resulting bands are shifted away from the Fermi level.

\begin{figure}[t!]
\begin{center}
\includegraphics[scale=1.8]{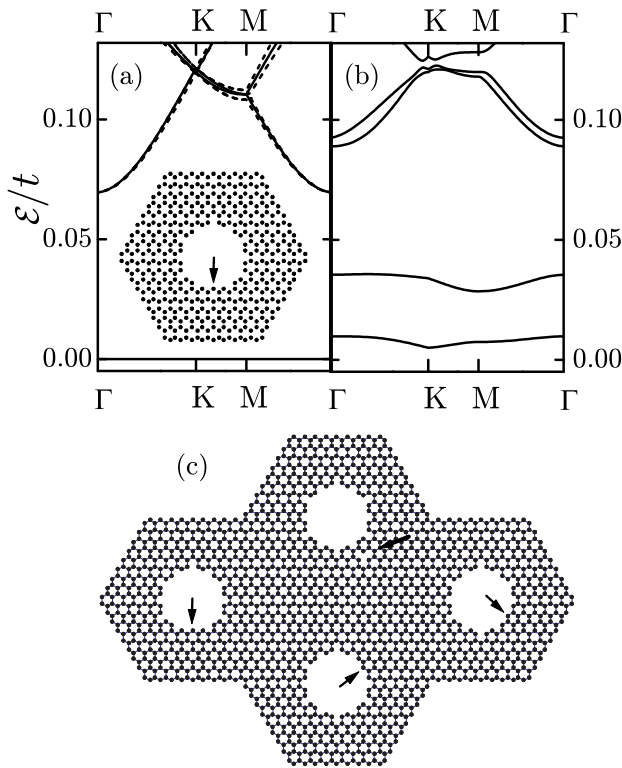}
\caption{\label{fig:WeakDisorder}
Effects of disorder in antidot lattices:
(a) band structure of an antidot lattice with defects
    on sublattice A (indicated by the arrow in the inset),
(b) band structure in the presence of defects on both sublattices,
and (c) unit cell with arrows indicating vacancies or C-adatom
defects. The dashed curves in (a) represent the band structure of
an ideal lattice (without disorder). $\Gamma$, K, M stand for the
high-symmetry points in the Brillouin zone of the antidot lattice with
the fourfold-enlarged unit cell.}
\end{center}
\end{figure}

In disordered antidot lattices, debonded C atoms generically appear along the edges of
irregularly shaped antidots, leading to a local sublattice imbalance. Such defects
at the edge of a zigzag nanoribbon have been experimentally
observed quite recently.~\cite{Liu+Iijima:09}
In this experiment, the multilayer graphene structures were thermally-treated leading to
edges that are mostly closed (i.e., folded from one layer to another). Open-edge structures with
debonded C atoms are found in a local area where the folding edge is partially broken.
Another source of local sublattice imbalances in disordered lattices are vacancies, for
example, due to removed C atoms or C atoms that are $sp^{3}$ rehybridized as a result of hydrogen
chemisorption.~\cite{Mizes+Foster:89,Ruffieux+:00,Rutter+:07} In Fig.~\ref{fig:WeakDisorder}(a),
we show that a single defect in an otherwise perfectly circular antidot lattice produces,
as expected, a zero-energy band. Needless to say, in realistic disordered systems defects
do not repeat with the lattice period as assumed in Fig.~\ref{fig:WeakDisorder}(a).
Some qualitative features of  realistic disordered lattices can be captured by
considering a superlattice with an increased unit cell. In
Fig.~\ref{fig:WeakDisorder}(b), we show a band structure for an
antidot lattice with a unit cell containing four antidots with
defects at different locations; this band structure indicates that
well-separated defect states weakly hybridize, giving rise to two
pairs of quasi-flat bands that are symmetric with respect to the
Fermi level. This suggests that for realistic disordered lattices
one has a quasicontinuum of such low-energy states due to hybridization
of defect states.

\begin{figure}
\begin{center}
\includegraphics[scale=2.05]{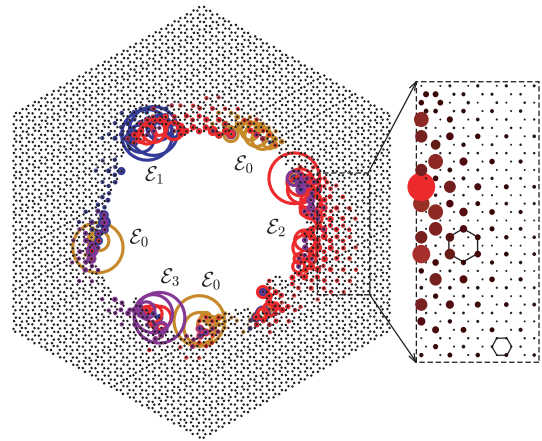}
\caption{\label{fig:IofRIrregularShape}
Tunneling current distributions for a lattice with
irregularly-shaped antidots, shown for the four lowest
electron bands: $\cE_{0}=0$ (brown), $\cE_{1}(\mathbf{k})$ (blue),
$\cE_{2}(\mathbf{k})$ (red), and $\cE_{3}(\mathbf{k})$ (violet).
The tunneling current magnitudes are represented
by circles of different sizes. The sublattice polarization due to
edge defects, along with the accompanying
charge-density reconstruction, is shown in the
zoomed part of the figure.}
\end{center}
\end{figure}

We finally consider an antidot lattice with strong shape irregularities, as
is likely the case in current experiments.~\cite{Shen:08pre,Eroms:09pre}
Although our calculation, the result of which is shown in Fig.~\ref{fig:IofRIrregularShape},
is based on periodically repeated defects, we expect that it will give reliable estimates
for the localization properties of the appearing
midgap states. The localization length found in this manner can be used to estimate the
charging energy in the system. The obtained band structure has the expected features:
the number of  flat bands at $\cE=0$ is in agreement with general result for the
bipartite systems, and the quasi-flat bands
at nonzero energies correspond to local sublattice imbalances. The electron states for both types
of flat bands are localized in the vicinity of the zigzag-like segments of the antidot edges
and debonded C atoms, as illustrated in Fig.~\ref{fig:IofRIrregularShape}.
(In contrast to the zigzag-like segments, their armchair-like counterparts on the antidot
edges do not exhibit electron localization.) Moreover, the tunneling current
is maximal at the midpoints of the zigzag-like segments on the antidot edges.
These features are in agreement with the existing experimental results on stepped
graphite surfaces,~\cite{Kobayashi+Kusakabe:06} which have edges with
similar irregularities.

Also depicted in Fig.~\ref{fig:IofRIrregularShape} (see the zoomed part) is a ``charge-density reconstruction''
from the original graphene honeycomb lattice into a lattice with $\sqrt{3}$ times larger period, which is rotated
by an angle of $30^\circ$. This is a manifestation of a Friedel-oscillation-like phenomenon,
that is, a long-range electronic perturbation caused by the presence of defects. Such interference phenomena, which
are essentially a consequence of the wave nature of electrons, are familiar in the
context of impurities on surfaces of metals~\cite{Hasegawa:93} and were  observed on the graphite
and graphene surfaces using STM.~\cite{Mizes+Foster:89,Mallet+:07} In the case of graphene, the
charge-density reconstruction results from intervalley coupling of the electronic $\pi$-states.
We thus conclude that there is strong intervalley scattering at the antidot edges in the  lattice shown
in Fig.~\ref{fig:IofRIrregularShape}. This is in agreement with the experiments of
Refs.~\onlinecite{Shen:08pre} and \onlinecite{Eroms:09pre}, where the observed weak localization (instead of weak antilocalization)
is a signature of intervalley scattering at the antidot edges.

The numerically obtained localization of the induced midgap states at the antidot
edges, with a localization length on the order of a few interatomic distances,
suggests that the charging energies for these states can be substantial.
Assuming completely random edges of length $L_{e}$,  one expects $|n_A-n_B|\sim\sqrt{L_{e}/a}$
low-energy states per unit cell due to local sublattice imbalances (in a realistic disordered
system these states will not be strictly at zero energy since the local sublattice imbalances
$n_A-n_B$ in individual unit cells will partially cancel between unit cells).
For the parameters in the experiment of Ref.~\onlinecite{Eroms:09pre}, one thus expects a spacing between
these localized states of about $\sqrt{L_e a}\sim 10\:{\rm nm}$ along the edge, implying a charging energy
that is substantially larger than the band gap in the system, which is roughly inversely proportional to the
distance between antidots. Also, since the low-energy edge states due to local sublattice imbalances are separated by the
antidot distance, one expects that their energy after hybridization is at most on the order of the band gap.
Figure \ref{fig:WeakDisorder}
supports this estimate. We thus conclude that the localized states observed in Fig.~\ref{fig:IofRIrregularShape}
have kinetic energies inside the band gap of the perfectly regular lattice, but charging energies that by far
exceed that band gap. This suggests that local spin moments may form at the  edges of disordered antidot
lattices at Fermi energies where charge transport takes
place. Since local magnetic moments are known to be an effective source of electron dephasing
in weak localization experiments,~\cite{dephasing} the studied midgap states offer an alternative
explanation for the saturation of the electron dephasing length at a scale corresponding to the distance
between antidots reported in Ref.~\onlinecite{Eroms:09pre}.

%For the sake of completeness, we emphasize some interesting aspects of
%flat-band physics in interacting systems. In such systems, due to vanishing
%kinetic energy, the effects of interactions are always non-perturbative.
%Magnetism in graphene-based systems was recently studied,~\cite{RossierPalacios,Yazyev:08}
%revealing that the existence of midgap states at zero energy is the physical origin
%behind both the vacancy-induced magnetism in graphene and that found in graphene flakes.
%Lieb's results~\cite{Lieb:89} were shown to be valid beyond the tight-binding approximation
%and the on-site Hubbard model. On the other hand, the flat bands that -- unlike the ones
%discussed in the present work -- arise as a result of interference between
%nearest-neighbor- and more distant electron transfers, play an important role in the
%Mielke-Tasaki mechanism of ``flat-band ferromagnetism''.~\cite{Mielke+Tasaki}
%This mechanism is relevant in some organic polymers~\cite{FlatBandFerroJPN} and semiconductor
%quantum-dot superlattices.~\cite{Tamura:02} Interesting effects have also been found in frustrated
%hopping models with flat bands,~\cite{Bergman+:08} such as the ones realized with cold-atoms
%(either bosons or fermions) in the $p$-orbital states of honeycomb optical
%lattices. The lowest energy band of this system is completely flat over the entire
%Brillouin zone.~\cite{Wu+Bergman:07}

\vspace*{5mm}
\section{Conclusions}\label{concl}

In this work, we have studied the salient features of
the low-energy band structure of graphene antidot lattices.
Apart from strictly zero-energy (midgap) flat bands that arise
from the bipartite lattice structure and a
global sublattice imbalance, we have also found
quasi-flat bands at low, but nonzero, energies that can be ascribed to
local sublattice imbalances. In addition, we have
examined the influence of an edge potential on the flat bands,
showing that such a potential lifts the degeneracy of these bands,
without affecting significantly their dispersionless character.

We have also investigated the spatial profile of the electronic states
corresponding to both classes of  low-energy bands (flat and quasi-flat).
By analyzing the tunneling-current distributions that can be compared
to STM measurements, we have demonstrated that these electronic states are
generically localized at the antidot-edges. The computed tunneling current
distributions also show a charge-density reconstruction from the
original honeycomb lattice to a lattice with $\sqrt{3}$ times larger
period and rotated through an angle of $30^\circ$. This phenomenon is indicative of
intervalley scattering off irregular antidot edges, as also observed
in recent experiments.~\cite{Shen:08pre,Eroms:09pre}

The spatial profiles of the localized midgap states that we find
allow for a rough estimate of their charging energies. That
estimate suggests that such states can host local magnetic
moments. We propose that such magnetic moments may be at the
origin of a recently observed saturation of the dephasing length
in graphene antidot lattices.~\cite{Eroms:09pre} In addition, the
investigated midgap states compete with the localized states due
to superlattice-scale defects, and therefore can have significant
implications for a recent proposal of spin qubits in graphene
antidot lattices.\cite{PedersenAntidot:08}

%{\em Note added}: Recent work (J. A. F\"{u}rst et al., arXiv:0904.1396)
%corroborates some of the results of the present paper by means of density functional
%theory calculations.

\section*{Acknowledgements}

We acknowledge a helpful correspondence with J. Eroms and
discussions with C. Flindt.

%\bibliography{Graphene}
\bibliographystyle{apsrev}

\end{document}